\documentclass[twocolumn, pra,superscriptaddress]{revtex4}
\usepackage{graphicx}
\usepackage{dcolumn}
\usepackage{bm}
\usepackage{amsmath}

\begin{document}

\preprint{APS/123-QED}

\title{Quench Dynamics of Anyon Tonks-Girardeau Gases}
\author{Yajiang Hao}
\email{haoyj@ustb.edu.cn}
\affiliation{Institute of Theoretical Physics and Department of Physics, University of Science and Technology Beijing, Beijing 100083, China}
\date{\today}

\begin{abstract}
We investigate the dynamical evolution of strongly interacting anyons confined in a weak harmonic trap using the exact anyon-fermion mapping method. The density profiles, momentum distribution, and the reduced one-body density matrix are obtained for different statistical parameters. The density profiles of anyons  display the same behaviors irrespective of statistical parameter during the evolution. As the harmonic trap is turned off suddenly, the momentum distributions exhibit the symmetric fermion-like behaviour in the long time evolution. As the trap frequency is quenched, the momentum distribution exhibit an asymmetry breath mode during the evolution. The reduced one-body density matrix show the dynamical symmetry broken and reproduced behaviour.
\end{abstract}
\maketitle


\narrowtext
\section{introduction}

In the last decades the dramatic experiment progress in the controlling optically trapped ultracold atomic gases stimulated the interests to study the nonequilibrium dynamics of isolated \cite{APolkovnikov} and open many-body quantum systems \cite{LMSieberer, HRitsch}. The high controllability of interaction \cite{BParedes,TKinoshita,TJacqmin,FR,CIR} and dimensionality \cite{NJVDruten,BParedes,TKinoshita} opens the way to investigate the fundamental questions concerning transport properties and relaxation dynamics of strong correlated system \cite{JPBrantut}. The ultracold atom system has become a popular platform to theoretically and experimentally investigate the nonequilibrium dynamics of quantum many-body systems at large distance and in long timescales \cite{RMP2011,RMP2012,RMP2013}. Besides the tunability of interaction and dimension, the tuning of quantum statistics \cite{anyons,Wilczek} is also feasible. The cold atom system has become the irreplacable platform for the research of interplay of interaction, dimension, and statistics in quantum many body system.

The strong correlated one-dimensional (1D) quantum gas is one of the research focuses in both experiment and theory because of its significant quantum effect and integrability \cite{RMP2011,RMP2012,RMP2013}. The combination of the highly controllable experiment technique and the exact solution of 1D many body model help us to understand the quantum effect at the untouchable level in the previous research. Theoretically the interacting 1D quantum system was proposed as 'toy' model at the earliest time such as the famous Tonks-Girardeau (TG) gas\cite{LTonks,MDGirardeau1960} that describe the Bose gas with infinite strong repulsive interaction. One of the most important quantum models is Lieb-Liniger model that describe the interacting bosons with contact interaction. The exact time-(in)dependent many-body wavefunction of TG gas can be obtained by Bose-Fermi mapping method. Using Bethe ansatz the ground state and thermodynamic properties of Lieb-Liniger gas in the whole interacting regime can be obtained. Comparing with the preceding theoretical research the 1D Bose gas is realized several decades years later \cite{BParedes,TKinoshita,NJVDruten}, but the realization induce the development of new theory, for example the application of generalized hydrodynamics theory to investigate the nonequalibrium dynamics of nonintegrable quantum many-body system. It is the mutual promotion and common development of theory and experiment that provide us more opportunity to understand the basic principles governing the nonequilibrium dynamics of many-body systems.

Not only the 1D Bose gas, the 1D Fermi gas, and quantum gas mixture have been realized, but also the anyon gas was paid great attentions. Anyons satisfy the fractional statistics \cite{Wilczek,anyons}, an intermediate one between Bose statistics and Fermi statistics, and have played important roles in condensed matter physics\cite{Wilczek,Laughlin,Halperin,Camino,YSWu,ZNCHa} including the explanation of fractional quantum Hall effect \cite{SuperConductor}. For the topological protection of quantum coherence \cite{RMP2008,Kitaev} the system satisfying fractional statistics has great potential application in quantum information science. Therefore the proposals to realize anyons in low-dimensional cold atom system have been suggested such as the schemes basing on the rotated Bose-Einstein condensates (BECs) \cite{PZoller}, the Raman-assisted hopping technique \cite{NatureComm,LSantos}, the lattice-shaking-induced tunneling \cite{Strater},
multicolor lattice-depth modulation \cite{LCardarelli,SGreschner}, and density-dependent gauge field \cite{LWClark}, etc.

Theoretically 1D anyon gas attracted many research interests such as the exact solution \cite{Kundu99,Girardeau06,Batchelor}, the low-energy properties \cite{XWGuanLowEnergy}, correlation function \cite{Patu07,Patu08,Calabrese,Cabra,anyonTG,HaoPRA78,HaoPRA79,YZhang}, entanglement properties \cite{Cabra,HLGuo}, the fermionization \cite{HaoPRA2012}, and anyon mixture \cite{Zinner}. Besides the static properties, the dynamical bosonization and fermionization \cite{Campo}, the relaxation dynamics in optical lattice \cite{MRigol}, interaction quench dynamics \cite{LPiroli} and the nonequilibrium dynamics at finite temperature \cite{OIPatu2020} are also investigated. In the present paper, we will study the relax dynamics of ground state of anyon TG gas induced by the quench of trap frequency including the sudden close of harmonic potential and the sudden change of trap frequency. Both the momentum distribution and the reduced one-body density matrix (ROBDM) in the nonequilibrium dynamics will be exhibited.

The paper is organized as follows. In Sec. II, we give a brief review of 1D anyonic model and introduce the analytical solution. In Sec. III, we present the evolution of the ROBDM and the momentum distributions. A brief summary is given in Sec. IV.

\section{model and method}
We investigate the evolution of $N$ anyons of mass $m$ with the infinite repulsive interaction, i.e., the anyon TG gases, trapped in a time-dependent harmonic potential
\[
V_{ext}(x,t)=m\omega^{2}\left(t\right)x^{2}/2.
\]
For anyonic system the wavefunction of $N$ anyons satisfy the generalized exchange symmetry \cite{Kundu99,AFM,HLGuo}
\begin{eqnarray}
&\Phi(x_{1},\cdots,x_j,\cdots,x_k,\cdots,x_{N};t) \nonumber    \\
&=e^{-i\theta}\Phi(x_{1},\cdots,x_k,\cdots,x_j,\cdots,x_{N};t)
\end{eqnarray}
with the anyonic phase
\begin{eqnarray*}
\theta=\chi\pi\left[ \sum_{l=j+1}^k\epsilon(x_j-x_l)-\sum_{l=j+1}^{k-1}\epsilon(x_k-x_l)\right]
\end{eqnarray*}
for $j<k$. The sign function $\epsilon(x)$ is 1, -1 and 0 for $x>0$, $<0$ and $0$, respectively. For the infinite repulsive interacting anyon gas we can construct the exact wavefunction basing on the wavefunction of the polarized free Fermi gas with the anyon-fermion mapping method \cite{AFM}
\begin{eqnarray}
&&\Phi(x_{1},\cdots,x_{N};t)		\\	\nonumber
&&=\mathcal{A}(x_{1},\cdots,x_{N})\Phi_{F}\left(x_{1},x_{2},\cdots,x_{N};t\right)
\end{eqnarray}
with the anyonic mapping function
\begin{equation}
\mathcal{A}(x_{1},\cdots,x_{N})=\prod_{1\leq j<k\leq N}\exp[-\frac{i\chi \pi}{2}\epsilon(x_j-x_k)].
\end{equation}
Here the phase factor is related to the quantum statistical property of anyons and the statistical parameter $\chi$ belongs to the interval $[0,1]$. The number 1 corresponds to the strongly interacting Bose gas and 0 corresponds to the polarized free Fermi gas. The time-dependent wavefunction of $N$ polarized fermions $\Phi_{F}\left(x_{1},x_{2},\cdots,x_{N};t\right)$ is the Slater determinant composed of the lowest $N$ eigenstates $\phi _j(x,t)$ ($j=1,\cdots,N$) of the single particle in the potential $V_{ext}(x,t)$ \cite{AMinguzzi}
\begin{equation}
\Phi_{F}\left(x_{1},x_{2},\cdots,x_{N};t\right)=\left(1/\sqrt{N!}\right)\det_{j,k=1}^{N}\phi_{j}\left(x_{k},t\right).
\end{equation}
The $j$th time-dependent eigenstate of eigenenergy $E_j$ take the  formulation of \cite{AMinguzzi,MCollura}
\begin{equation}
\phi_{j}\left(x,t\right)=\frac{1}{\sqrt{b}}\phi_{j}(\frac{x}{b},0)\exp[i\frac{mx^{2}}{2\hbar}\frac{\dot{b}}{b}-iE_{j}\tau(t)],
\end{equation}
where the coordinate and time has been rescaled by the scaling factor $b(t)$ that satisfies the differential equation
\begin{equation}
\ddot{b}+\omega^{2}(t)b=\omega_{0}^{2}/b^{3}
\end{equation}
with the initial condition $b(0)=1$ and $\dot{b}(0)=0$. For the free expansion case, i.e., the harmonic potential is turned off at $t=0$ ($\omega(t)=\omega_0$ for $t\leq0$ and $\omega (t)=0$ for $t>0$) the scaling factor is expressed as $b(t)=\sqrt{1+\omega_0^2t^2}$. For the trap frequency quench ($\omega(t)=\omega_0$ for $t\leq0$ and $\omega (t)=\omega_1$ for $t>0$) the scaling factor is expressed as $b(t)=\sqrt{1+(\omega_0^2-\omega_1^2)\sin^2(\omega_1t)/\omega_1^2}$. The time is rescaled as $\tau(t)=\int_{0}^{t}dt'/b^{2}(t')$. The initial state $\phi _j(x,0)$ at $t=0$ is the $j$th eigenstate of harmonic oscillator with frequency $\omega (0)$. Replacing Eq. (3) and Eq. (4) into Eq.(1), we have the exact time-dependent anyon wavefunction
\begin{eqnarray}
&&\Phi(x_{1},\cdots,x_{N};t)	\\	\nonumber
& =& b^{-N/2}\Phi(x_{1}/b,\cdots,x_{N}/b;0)\exp[\frac{i\dot{b}}{b\omega_{0}}\sum_{l}\frac{x_{l}^{2}}{2l_{0}^{2}}]	\\	\nonumber
&&\times \exp[-i\sum_{l}E_{l}\tau(t)].
\end{eqnarray}

The ROBDM can be evaluated with the above time-dependent wavefunction and formulated as
\begin{align}	\nonumber
\rho(x,y,t) & =N\int dx_{2}\cdots dx_{N-1}\Phi^{*}(x_{1},\cdots,x_{N-1},x;t) \nonumber   \\ &\times \Phi(x_{1},\cdots,x_{N-1},y;t)    \nonumber\\
 & =\frac{1}{b}\rho(\frac{x}{b},\frac{y}{b},0)\exp[-\frac{i\dot{b}}{b\omega_{0}}\frac{x^{2}-y^{2}}{2l_{0}^{2}}].	
\end{align}
Here $l_0=\sqrt{\hbar/m\omega_0}$ is the typical length of harmonic potential with the initial trap frequency. During the evolution besides the time-dependent phase factor the initial ROBDM is rescaled by the scaling factor $b$. It is interesting to notice that the expression Eq. (8) is same as the Bose case \cite{AMinguzzi}. The statistical properties of anyons will depend only on the rescaled initial ROBDM. The time-dependent phase factor of the anyon TG gases is exactly same as those of the Bose limit and that of the free Fermi limit. This is because the time-dependent phase factor in wavefunction Eq. (7) is independent on the statistical parameter. With the properties of Vandermonde determinant and Hankel type determinant \cite{PJForrester,Hao2016} the concise expression of the initial ROBDM $\rho (x,y,0)$ can be formulated as \cite{Hao2016}.
\begin{eqnarray}
\rho\left(x,y,0\right)&=&\frac{2^{N-1}}{\sqrt{\pi}\Gamma\left(N\right)}exp(-x^{2}/2-y^{2}/2)	\\	\nonumber
&& \times \det[\frac{2^{(j+k)/2}}{2\sqrt{\pi}\sqrt{\Gamma\left(j\right)\Gamma\left(k\right)}}\beta_{j,k}\left(x,y\right)]_{j,k=1,\cdots,N-1}
\end{eqnarray}
with
\begin{eqnarray}
&&\beta_{j,k}(x,y)=		\\	\nonumber
&&f_{j,k}(x,y)+(e^{i\chi\pi\epsilon(y-x)}-1)\epsilon(y-x) [xy	\\	\nonumber
&&\times \mu_{j+k-2}(x,y) -(x+y)\mu_{j+k-1}(x,y)+\mu_{j+k}(x,y)].
\end{eqnarray}
Here the function $f_{j,k}(x,y)$ depends on Gamma function and $\mu_m(x,y)$ depends on the confluent hypergeometric function \cite{PJForrester}.

The diagonal part of ROBDM is the density profile $\rho(x,t)=\rho(x,x,t)$ in coordinate space while the momentum distribution is evaluated by the Fourier transformation of ROBDM. If we rescale the coordinate as a time-dependent quantity, the momentum distribution can be formulated as
\begin{eqnarray}
 n(k,t)&=&\frac{b}{2\pi}\int dx'dy'\rho(x',y',0)	\\	\nonumber
&&\times \exp[-ib(\frac{\dot{b}}{\omega_{0}}\frac{x'^{2}-y'^{2}}{2l_{0}^{2}}-\frac{k(x'-y')}{\hbar}]
\end{eqnarray}
with $x'=\frac{x}{b}$ and $y'=\frac{y}{b}$.

\section{numerical result}

\begin{figure}
\includegraphics[width=3in]{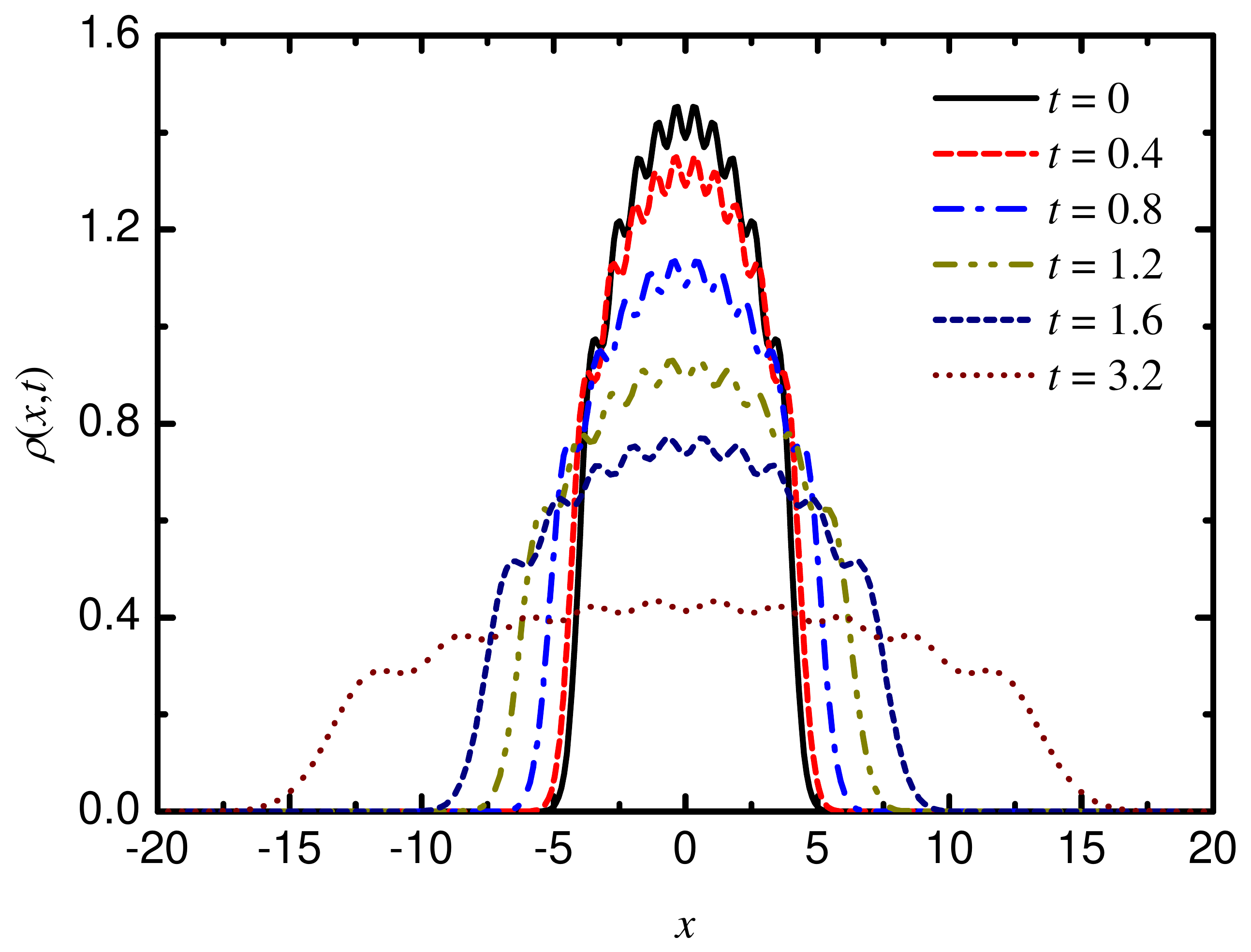}
\caption{ The free expansion of anyon TG gas with $N=10$ after the harmonic potential is turned of at $t=0$. $x$ is in unit of $\sqrt{\hbar/m\omega_0}$; The time $t$ is in unit of $1/\omega_0$.}
\end{figure}

For convenience in the following we take the length unit as $\sqrt{\hbar/m\omega_0}$ and time unit as $1/\omega_0$ and the original notation will be preserved.

The density distribution of anyon TG gas is independent on the the statistical parameter $\chi$. In Fig. 1 we display the free expansion of anyon TG gas with $N=10$. It is shown that with the time evolution the density distribution keeps expanding and anyons distribute in larger regime after the harmonic potential is turned off suddenly. During  the expansion the density profile always display the typical shell structure of TG gases and the peak number is equal to the particle number ($N=10$ in the present calculation).

\begin{figure}
\includegraphics[width=3.4in]{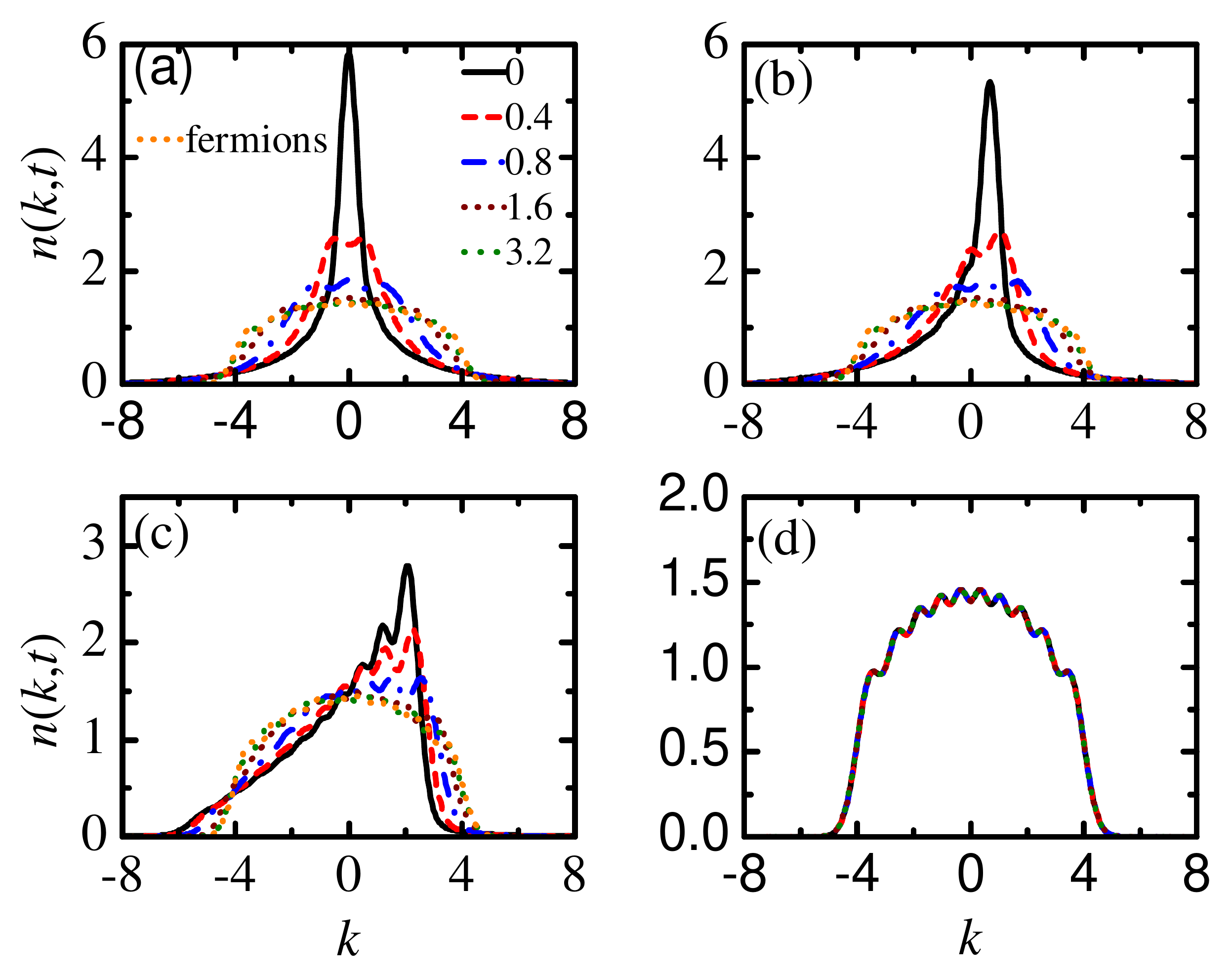}
\caption{The evolving momentum distribution at different time for free expanded anyon TG gas with $N=10$. (a) $\chi$=1.0; (b) $\chi$=0.8; (c) $\chi$=0.4; and (d) $\chi$=0.0. As a comparison, the initial momentum distribution and density profiles $\rho(x,0)$ of free fermions is plotted in (a)-(c). $k$ is in unit of $\sqrt{m\omega_0/\hbar}$; The time is in unit of $1/\omega_0$.}
\end{figure}
\begin{figure*}
\includegraphics[width=7in]{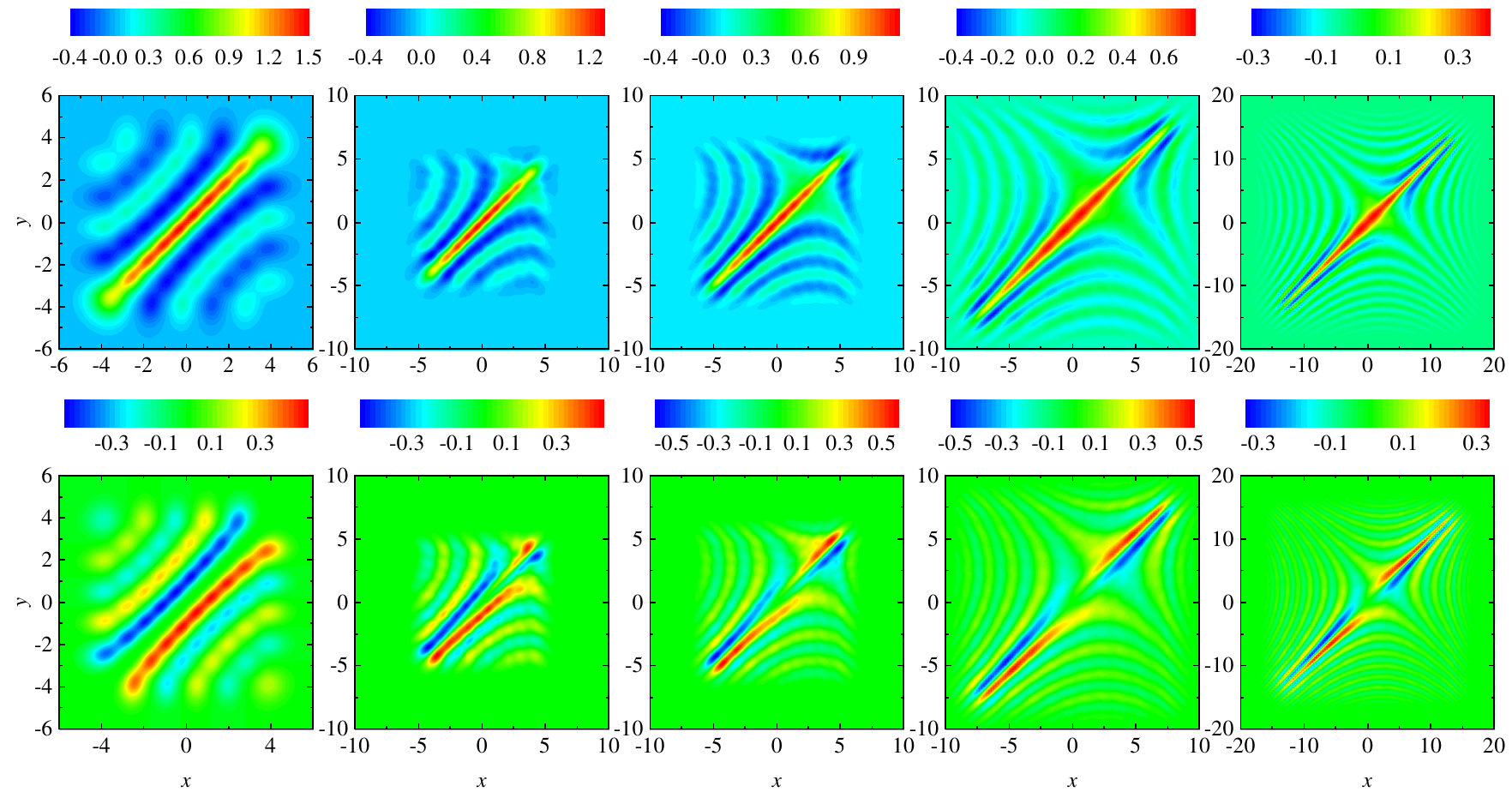}
\caption{The ROBDM of Anyons of $\chi = 0.5$ for $N$=10 after the potential is turned off. The first row: Real part; The second row: Imaginary part. From left to right: $t$=0, 0.4, 0.8, 1.6 and 3.2. $x$ and $y$ are in units of $\sqrt{\hbar/m\omega_0}$; The time $t$ is in units of $1/\omega_0$.}
\end{figure*}
\begin{figure}
\includegraphics[width=3in]{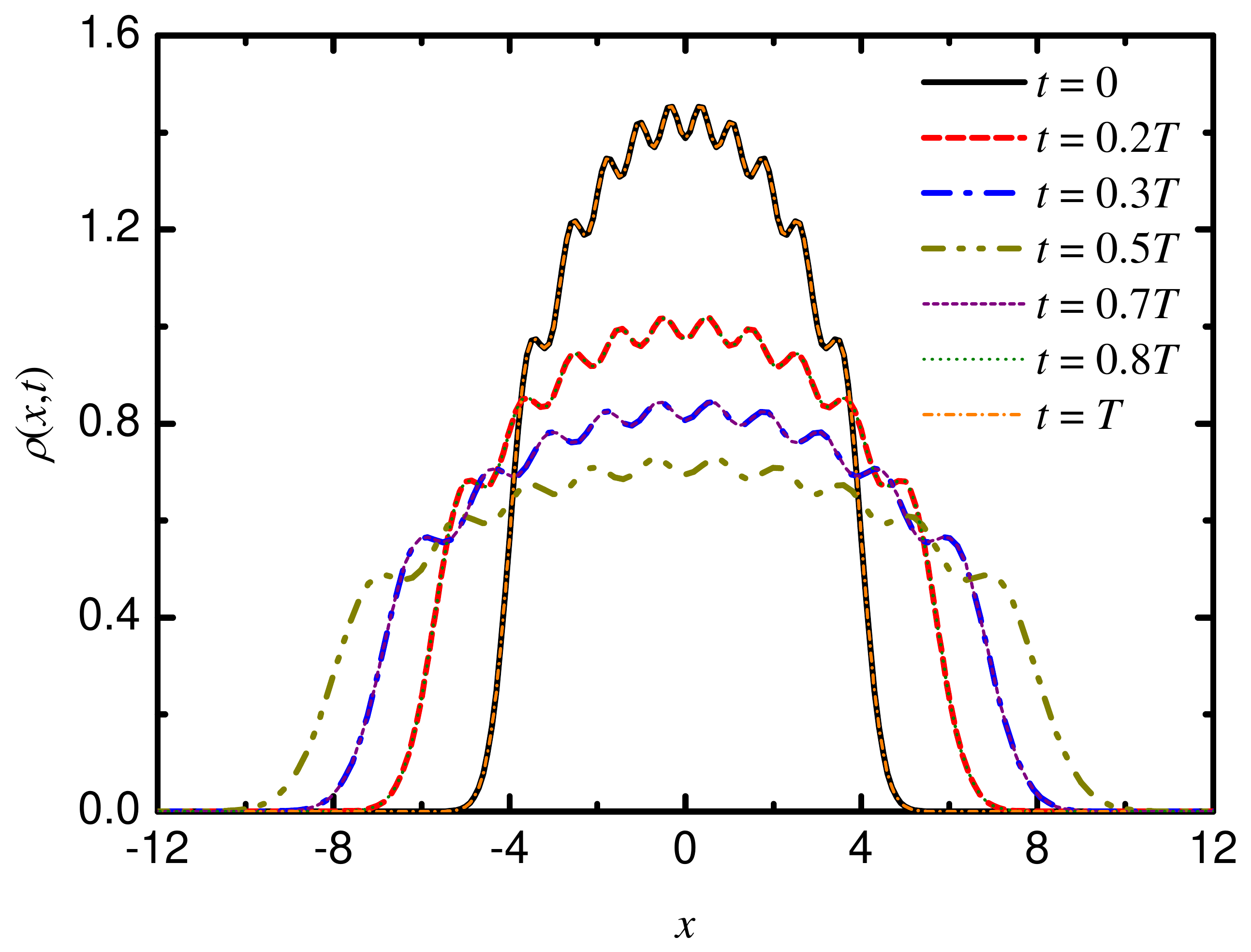}
\caption{Density profiles evolving in one period for 10  Anyons of $\chi = 0.5$ in a weaker harmonic trap. $\omega_0$=1.0 and $\omega_1$=0.5. $x$ is in unit of $\sqrt{\hbar/m\omega_0}$; The time $t$ is in unit of $1/\omega_0$.}
\end{figure}
\begin{figure}
\includegraphics[width=3.4in]{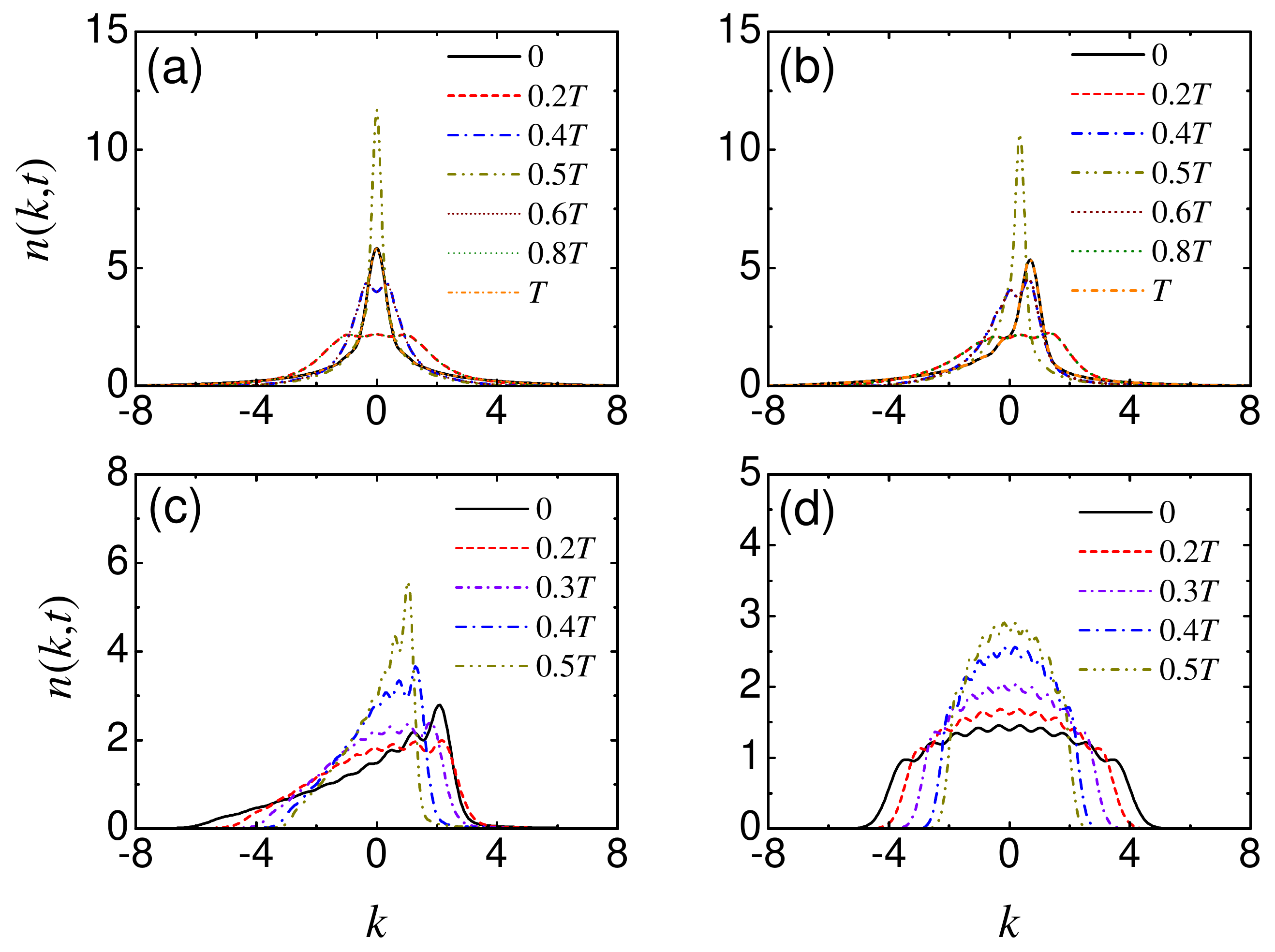}
\caption{Momentum distribution evolution for 10  Anyons of $\chi = 0.5$ in a weaker harmonic trap. $\omega_0$=1.0 and $\omega_1$=0.5. (a) $\chi$=1.0; (b) $\chi$=0.8; (c) $\chi$=0.4; and (d) $\chi$=0.0. $k$ is in unit of $\sqrt{m\omega_0/\hbar}$; The time is in unit of $1/\omega_0$.}
\end{figure}
\begin{figure*}
\includegraphics[width=7in]{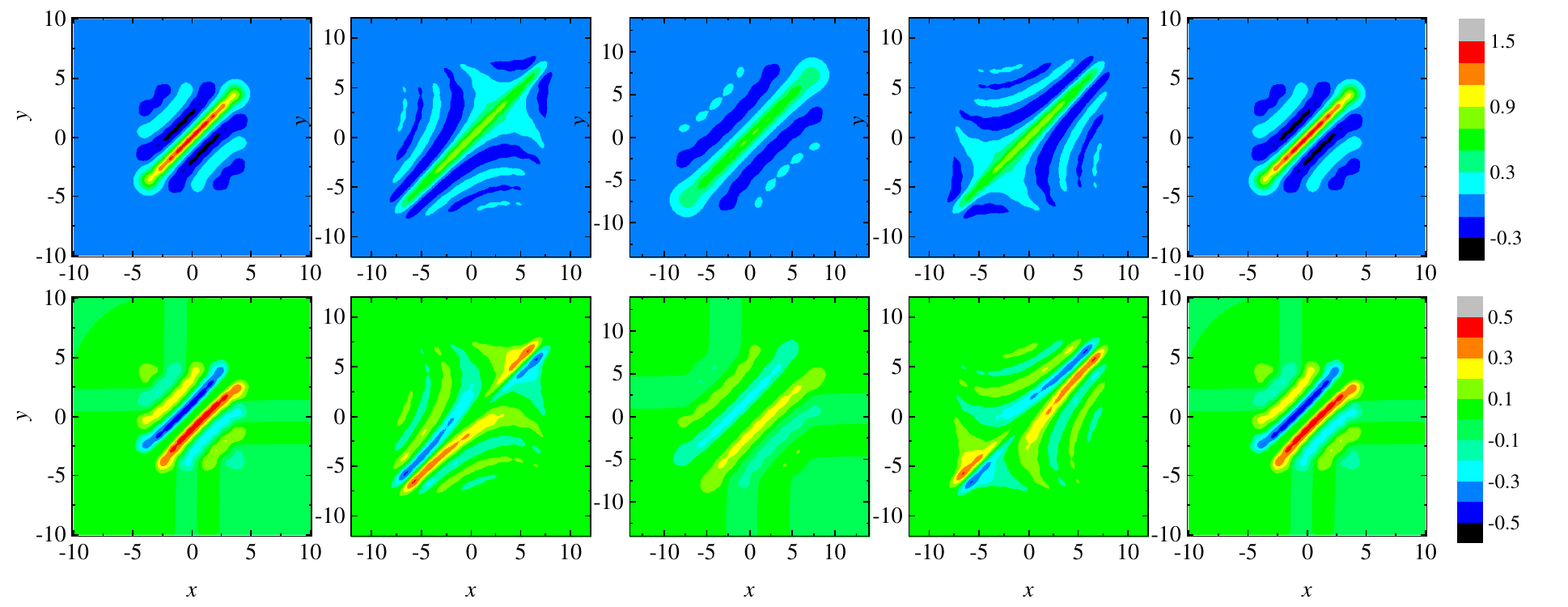}
\caption{The ROBDM of Anyons of $\chi = 0.6$ for $N$=10 evolved in one period. The first row: Real part; The second row: Imaginary part. From left to right: $t$=0, 0.3$T$, 0.5$T$, 0.7$T$ and $T$. $x$ and $y$ are in units of $\sqrt{\hbar/m\omega_0}$; The time $t$ is in units of $1/\omega_0$.}
\label{default}
\end{figure*}

The evolving momentum distributions of anyon TG gases are displayed in Fig. 2 for different statistical parameters. The strongly interacting Bose limit and the free Fermi limit are displayed in Fig. 2a ($\chi=1.0$) and in Fig. 2d ($\chi=0.0$), respectively. Initially the strongly interacting Bosons show the typical symmetric single-peak structure and the Bose gas distribute in the low momentum regime with larger probability. After the harmonic potential is turned off Bosons expand into large momentum regime and more and more peaks appear. The initial single-peak structure gradually evolve into multiple-peak structure. As the evolution time is long enough the TG gases display the same distribution as that of polarized free Fermions. As the comparison, we plot the momentum distribution of free fermions in Fig. 2(a-c) by the dot lines. In the limit of free Fermions it is interesting that the momentum distributions do not change at all in the momentum space although they expand freely in the coordinate space. The intermediate case, i.e., the evolving momentum distributions of anyon TG gases are displayed in Fig. 2b ($\chi=0.8$) and Fig. 2c ($\chi=0.4$). At the initial time the asymmetric momentum distributions are displayed for both cases. The former show single-peak structure and the later show multiple-peak structure with different peak height. As the time goes on the anyon TG gases evolve from the asymmetric momentum distribution into the symmetric profiles and finally exhibit the same behaviour as those of free fermions. It is also interesting to notice that in coordinate space the anyon TG gases keep expanding but in momentum space the momentum distributions evolve in a finite regime.

The ROBDM of the anyon TG gas is a complex conjugate matrix. In Fig. 3 we display the evolution of the ROBDM of the anyon TG gas with statistical parameter $\chi=0.5$. The evolution of the real part is plotted in the first row, which is symmetry about $y=x$ and $y=-x$ at the initial time, and the evolution of imaginary part is plotted in the second row, which is antisymmetry about $y=x$ but is symmetry about $y=-x$ at the initial time. During the evolution the symmetry of the real part about $y=x$ always preserves but its symmetry about $y=-x$ breaks first and finally reproduce at long enough time. The evolving real part is always diagonal dominated but the alternating stripes gradually shrink in narrower regime nearby the diagonal and change into the shape of shuttle with two sharp points. The imaginary part also exhibits the broken symmetry about $y=-x$ as time goes on and finally evolves into the anti-symmetry about $y=-x$. At long enough time the imaginary part shows the antisymmetry about both $y=x$ and $y=-x$.

Another case in experiments is the quench to a different trap frequency. According to the scaling factor it is obvious that after the potential quench the wavefunction, therefore the density profiles, the ROBDM and the momentum distribution evolve periodically with the period $T=\pi/\omega_1$. In Fig. 4 the evolution of density profiles in one period are plotted for the trap frequency $\omega_1=\omega_0/2$. It is shown that the anyon TG gases display typical breath mode with the period $T=2\pi/\omega_0$. In the first half period anyons expand in larger region while in the second half period anyons shrink back into the original region. Finally at time $T$ anyon TG gas display the same density profile as that at the initial time. The density profiles match each other at $t$ and at $T-t$.

The evolving momentum distribution for anyon TG gas with $\omega_1=\omega_0/2$ are displayed in Fig. 5 for different statistical parameter. In Fig. 5a (the Bose limit with $\chi=1.0$) and in Fig. 5b ($\chi=0.8$) the evolution of momentum distribution were plotted in the whole period which show that $n(k,t)$ are same as $n(k,T-t)$. This is true in all cases so in Fig. 5c and in Fig. 5d only the first half period are plotted.

In the strongly interacting Bose limit (Fig. 5a) and in the free Fermi limit (Fig. 5d) the momentum distributions are always symmetric during the evolution. In the Fermi limit the momentum distribution shrink continuously in the first half period. After the trap frequency quench to $\omega_0/2$ the momentum distribution of free fermions exhibit breath mode in which they keep shrinking in the first half period and keep expanding in the second half period until evolve back to the initial distribution. While in the Bose limit the momentum distribution do not keep shrinking or expanding in half period. It is contrary to the breath mode in the coordinate space where the density profiles expand first and then shrink back. The Bosons expand into multiple-peak structure at first and then shrink into a narrower single-peak structure and at half period the height of the momentum distribution arrive at the highest value that is higher than that at the initial time. In the second half period the opposite process are displayed. At $t=0.2T$ the structure with three peaks is shown and at $t=0.4T$ the structure with two peaks is shown. Fig. 5b ($\chi=0.8$) and Fig. 5c ($\chi=0.4$) show the asymmetric momentum distributions that are significantly different from that of Bosons and Fermions. In the later case the momentum distributions keep shrinking and the height keep increasing in the first half period, which is similar to the case of the Fermi limit. While in the former case the momentum distributions expand to multiple-peak structure and then shrink to higher single-peak structure and arrive at the largest height at $T/2$, which is similar to the case of the Bose limit. For the anyon TG gases the momentum distributions shrink in a narrow region at half period but they are asymmetric about $k=0$ and the peaks deviate from the zero momentum. In short, during the evolution the anyon TG gas display the symmetric breath mode in the coordinate space while the momentum distribution display the asymmetric breath mode.

In Fig. 6 we display the evolution in one period of the ROBDM of the anyon TG gas with statistical parameter $\chi=0.6$ as the trap frequency decrease one half suddenly. The real part of the ROBDM are plotted in the first row, and the imaginary part of the ROBDM are plotted in the second row. Same as before, the ROBDM at time $T$ show the same behaviour as the ROBDM shows at the initial time. During the evolution the ROBDM always preserve the Hermite in the whole period. But the symmetry about $y=-x$ breaks at first and then reproduce at $T/2$, and after $T/2$ the symmetry breaks again and then reproduces finally at $T$. During the evolution the real part of the ROBDM is always diagonal dominated but the difference between the diagonal and non-diagonal is not as larger as that at the initial time. The imaginary part also show the similar process. It is also interesting to notice that the ROBDM at $t$ and the ROBDM at $T-t$ are symmetric about $y=-x$. This is different from the equivalent of $n(k,t)=n(k,T-t)$.


\section{Summary}
In conclusion, with the anyon-fermion mapping method we obtained the exact time-dependent wavefunction of the anyon TG gases as the trap frequency quench to zero or to a different frequency. Basing on the wavefunction, we evaluated the evolution of density profiles, momentum distribution and the ROBDM for two typical experiment scenarios. It is shown that the many-body wavefunction and the density profiles exhibit the self-similar structure in the coordinate space during the evolution, but the momentum distribution and the ROBDM do not.

In the free expansion case the density profiles of the anyon TG gases are independent on the statistical parameter and keep expanding. The dependence on the statistical parameter is displayed in momentum distribution, which is symmetric about zero momentum in the strongly interacting Bose limit and free Fermi limit and is asymmetric in the intermediate case. In the free Fermi limit although the density profiles keep expanding but the momentum distribution always keep its initial profiles. In all other cases the momentum distribution evolve in a finite regime and finally display the typical behaviour of free Fermions at long enough time.

In the trap frequency quench case, the density profiles in coordinate space display the periodical symmetric breath mode and the momentum distributions display asymmetric breath mode. In the free Fermi limit the momentum distribution shrink into smaller region in the first half period and expand in the second half period, which is contrary to the breath mode of density profiles in the coordinate space. The momentum distribution for other statistical parameter do not keep shrinking or expanding in half period, which expand into a multiple peaks structure firstly and then shrink into a single higher peak structure at half period, then the reverse process evolve in the second half period.

During the evolution the symmetry of ROBDM about $y=x$ preserves but the symmetry about $y=-x$ breaks firstly and reproduces finally.

\begin{acknowledgments}
This work was supported by NSF of China under Grants No. 11174026.
\end{acknowledgments}



\end{document}